\documentclass[12pt]{article}
\usepackage[dvips]{graphicx}
\usepackage{epsfig}
\usepackage{cite}
\usepackage{latexsym,amsmath,amsfonts,amssymb}
\usepackage{mathtools} 
\usepackage{slashed} 
\usepackage{blkarray}
\usepackage[all]{xy} 
\usepackage[makeroom]{cancel}
\usepackage[latin1]{inputenc}
\usepackage[american]{babel}
\usepackage[dvips]{graphicx}
\usepackage{bbm}
\usepackage{color}
\usepackage[unicode]{hyperref}
\def\6{\partial}

\newcommand{\be}{\begin{equation}}
\newcommand{\ee}{\end{equation}}
\newcommand{\beq}{\begin{equation}}
\newcommand{\eeq}{\end{equation}}
\newcommand{\bea}{\begin{eqnarray}}
\newcommand{\eea}{\end{eqnarray}}

\newcommand{\ba}{\begin{eqnarray}}
\newcommand{\ea}{\end{eqnarray}}

\newcommand{\beqs}{\begin{eqnarray}}
\newcommand{\eeqs}{\end{eqnarray}}
\newcommand{\bal}{\begin{aligned}}
\newcommand{\eal}{\end{aligned}}

\def\lbldef#1#2{\expandafter\gdef\csname #1\endcsname {#2}}

\def\href#1#2{#2}

\newcommand{\ber}{\begin{eqnarray}}
\newcommand{\eer}{\end{eqnarray}}

\newcommand{\beqar}{\begin{eqnarray}}

\newcommand{\eeqar}{\end{eqnarray}}

\newcommand{\dsl}
   {\kern.06em\hbox{\raise.15ex\hbox{$/$}\kern-.56em\hbox{$\partial$}}}

\newcommand{\eeqarr}{\end{eqnarray}}
\newcommand{\ZZ}{{\rm \kern 0.275em Z \kern -0.92em Z}\;}

\def\CC{{\mathchoice
{\rm C\mkern-8mu\vrule height1.45ex depth-.05ex
width.05em\mkern9mu\kern-.05em}
{\rm C\mkern-8mu\vrule height1.45ex depth-.05ex
width.05em\mkern9mu\kern-.05em}
{\rm C\mkern-8mu\vrule height1ex depth-.07ex
width.035em\mkern9mu\kern-.035em}
{\rm C\mkern-8mu\vrule height.65ex depth-.1ex
width.025em\mkern8mu\kern-.025em}}}

\def\RR{{\rm I\kern-1.6pt {\rm R}}}

\def\ZZ{{\rm Z}\kern-3.8pt {\rm Z} \kern2pt}
\def\IB{\relax{\rm I\kern-.18em B}}
\def\ID{\relax{\rm I\kern-.18em D}}
\def\II{\relax{\rm I\kern-.18em I}}
\def\IP{\relax{\rm I\kern-.18em P}}

\newcommand{\bear}{\begin{eqnarray}}
\newcommand{\eear}{\end{eqnarray}}

\def\6{\partial}

\newfont{\namefont}{cmr10}
\newfont{\addfont}{cmti7 scaled 1440}
\newfont{\boldmathfont}{cmbx10}
\newfont{\headfontb}{cmbx10 scaled 1728}

\newcommand{\dd}{\textrm{d}}

\allowdisplaybreaks[1]

\numberwithin{equation}{section}

\begin{document}

\begin{titlepage}

\vfill
\begin{flushright}
\end{flushright}

\vfill

\begin{center}
   \baselineskip=16pt
   {\Large \bf {A hint of matter underdensity  at  low $z$?}}
   \vskip 2cm
   Eoin \'O Colg\'ain$^{a, b}$
          \vskip .6cm
             \begin{small}
               \textit{$^a$ Asia Pacific Center for Theoretical Physics, Postech, Pohang 37673, Korea}
               
               \vspace{3mm} 
               
               \textit{$^b$ Department of Physics, Postech, Pohang 37673, Korea}
  
             \end{small}
\end{center}

\vfill \begin{center} \textbf{Abstract}\end{center} \begin{quote}
{The $\Lambda$CDM cosmological model provides to first approximation a good description of the universe, but various tensions with data, most notably Hubble tension, persist. In this work we confront $\Lambda$CDM with the Pantheon Type Ia supernovae dataset and perform a two-parameter fit of the distance modulus for a running cut-off $z_{\textrm{max}}$. 
We observe that in a window between $z_{\textrm{max}} \approx 0.1$ and $z_{\textrm{max}} \approx 0.16$ there is a 1 - 2  $\sigma$ discrepancy with the Planck value $\omega_m = 0.315 \pm 0.007$, which points to a potential matter underdensity. For high-energy theorists, the analysis appears to support the de Sitter Swampland conjecture. }
\end{quote} \vfill

\end{titlepage}

\section{Introduction}
Ever since the seminal Riess, Macri et al. local determination of the Hubble constant $H_0$ to $2.4\%$ uncertainty \cite{Riess:2016jrr}, the tension with the Planck value based on $\Lambda$CDM \cite{Aghanim:2018eyx}, dubbed ``$H_0$ tension", has been difficult to ignore. Starting with a difference of $3.4 \, \sigma$, we have witnessed a steady growth in the discrepancy to the point that the statistical significance is now $4.4 \, \sigma$ \cite{Riess:2019cxk}. Despite the immense success of $\Lambda$CDM - built on the assumption that our universe is described by a cosmological constant and cold dark matter - this brings us potentially closer to a point in time when the standard model of cosmology may be due a slight tweak. That being said, in spite of the tension, the fact that measurements of $H_0$ based on radically different experiments at different redshifts agree so well is truly remarkable. 

In line with steadily more precise local determinations of $H_0$ over recent years, we have witnessed a migration in theory. Going beyond the assumption of a cosmological constant, there is now no shortage of dark energy models on the market \cite{Copeland:2006wr}. At one end of the spectrum (of speculation), one finds the de Sitter Swampland conjecture \cite{Obied:2018sgi}, which claims that de Sitter vacua belong to the ``Swampland" \cite{Vafa:2005ui} of inconsistent low-energy theories coupled to gravity, and for this reason, de Sitter vacua are ruled out. Bearing in mind that de Sitter is an attractor for $\Lambda$CDM, the conjecture is also in tension with $\Lambda$CDM. The de Sitter Swampland conjecture is controversial \cite{Garg:2018reu, Denef:2018etk, Andriot:2018wzk} (see \cite{Palti:2019pca} for a Swampland review), but the implication for $\Lambda$CDM appears clear. It has recently been explained \cite{Ooguri:2018wrx} how the conjecture can be motivated from the distance conjecture \cite{Ooguri:2006in} and Bousso covariant entropy bound \cite{Bousso:1999xy}, thus placing it on firmer theoretical footing: the conjecture may be here to stay \footnote{See \cite{Agrawal:2018own, Colgain:2018wgk, Heisenberg:2018yae, Heckman:2019dsj, Heisenberg:2019qxz, Brahma:2019kch} for dark energy implications of the conjecture.}. 

Taken together $H_0$ tension and the de Sitter Swampland suggest something is up with $\Lambda$CDM. Of course, the latter doesn't pinpoint a point in time where a deviation from $\Lambda$CDM is expected, but suggests something should happen before we meet the future asymptotic de Sitter attractor. On the contrary, the former is more concrete. There are a number of tensions between $\Lambda$CDM and existing datasets \cite{Heymans:2012gg, Troxel:2017xyo, Joudaki:2017zdt, Hikage:2018qbn}, but the most striking clash can be found in a local measurement of $H_0$  \cite{Riess:2019cxk}. Inspired by $H_0$ tension, we will perform an analysis of $\Lambda$CDM at low redshift $z$, which we will fit to the Pantheon compilation of Type Ia supernovae \cite{Scolnic:2017caz} (see \cite{Maurice} for an alternative perspective). Recall that Type Ia supernovae were instrumental in providing initial evidence for late time cosmic acceleration \cite{Riess:1998cb, Perlmutter:1998np} and have proven themselves to be robust probes of cosmological parameters. In contrast to other cosmological studies, e. g.  \cite{Zhao:2017cud, Dutta:2018vmq, Wang:2018fng}, here we will adopt a minimal approach and work within a single dataset. The Pantheon dataset consists of 1048 supernovae in the redshift range $0.01$ - $2.26$ and the idea is simply to impose a cut-off $z_{\textrm{max}}$ and restrict the analysis to supernovae below this value. 

Let us attempt to justify why this is an interesting exercise. First, at late times or low redshift, $\Lambda$CDM is expected to be described by an analytic solution to the Friedmann equation, 
\be
\label{LCDM}
H(z) = H_0 \sqrt{1 - \omega_{m} + \omega_{m} (1+z)^3}, 
\ee
where the Hubble constant $H_0$ and the matter density $\omega_m$ are the only free constant parameters. The expression is valid for a FLRW metric that is three-flat, so we have neglected spatial curvature. A connected observation is that $H_0$ is an overall factor that is insensitive to $z$ and can be determined at $z \approx 0$, so when Riess et al. determine $H_0$ they can do so in principle without assuming a cosmology. However, here the cosmology, namely $\Lambda$CDM, is captured by the constant $\omega_m$. Secondly, given that (\ref{LCDM}) is analytic, this means that if one expands at small $z$, one inevitably picks up information about $\omega_m$, so that one can start probing $\Lambda$CDM through the value of $\omega_m$. Of course, if the dataset is too sparse, there will be considerable uncertainty in the best-fit value of $\omega_m$, but Pantheon has 630, 832, and 1025 supernovae below $z=0.3$, $z = 0.5$ and $z =1$, respectively. Third, it is insightful to see how the best-fit value of the cosmological parameters changes if we only had access to supernovae below a given redshift.  On one hand, a good model is one where the parameters are robust to such changes, ideally within a $1 \, \sigma$ confidence window, and it is a valid exercise to check this. On the other, the JWST will extend the Hubble diagram to even higher redshifts, potentially $z \approx 5$ \cite{Scolnic:2019apa}, so local determinations of cosmological parameters based on supernovae may be expected to change as we access higher redshift data. 

Concretely, in this note we impose a series of cut-offs $z_{\textrm{max}}$ and identify the best-fit parameters $(H_0, \omega_m)$. Strictly speaking we cannot determine $H_0$ from Type Ia supernovae alone, since $H_0$ is degenerate with the absolute magnitude $M$. That being said, the latter is a constant and is not expected to depend on redshift, so we will just assume a nominal value for $M$ in performing fits \footnote{A proper determination of $H_0$ requires a knowledge of the absolute magnitude $M$ of Type Ia supernovae, and this is not possible without using Cepheids to break the degeneracy between $H_0$ and $M$. See \cite{Riess:2009pu, Riess:2011yx} for earlier studies in this direction.}. Therefore, since $\omega_m$ is the only term we can properly determine, we focus on it. We can summarise our findings succinctly. Although our ``$H_0$" is actually a combination of $H_0$ and $M$, we see that in line with expectations it is robust to changes in $z_{\textrm{max}}$. The same cannot be said for $\omega_m$. As can be anticipated for small $z_{\textrm{max}} < 0.1$, the best-fit values of $\omega_m$ are uncertain: the $1 \, \sigma$ confidence intervals are large and cover the Planck value $\omega_m = 0.315 \pm 0.007$. However, the key take-home message is that there exists an intermediate range of redshift {between $z_{\textrm{max}} \approx 0.1$ and $z_{\textrm{max}} \approx 0.16$}  where the confidence intervals contract and the best-fit value of $\omega_m$ exhibits a discrepancy with the Planck value that {approaches} $2 \, \sigma$. Extending the cut-off to higher redshift, the best-fit value converges to the quoted Pantheon value \cite{Scolnic:2017caz}, thus providing an important consistency check on our methods.

\section{Data Fitting}
In this section we introduce our assumptions and proceed to fit the $\Lambda$CDM model to the Pantheon supernovae data \cite{Scolnic:2017caz} for a running cut-off $z_{\textrm{max}}$. Our first input is that $\Lambda$CDM is described by only two parameters at late times through the Hubble parameter (\ref{LCDM}). By definition, the luminosity distance is 
\be
\label{ld}
d_{L}(z) = c (1+z) \int_{0}^{z} \frac{\dd z'}{H(z')}, 
\ee
where $c$ is the speed of light, and this serves as an input in the distance modulus
\bea
\label{distance_modulus}
\mu &=& m_{B} - M = 5 \log_{10} \left( \frac{d_{L}}{10 \textrm{pc}} \right) = 25 + 5 \log_{10} \left( \frac{ d_{L}}{\textrm{Mpc}} \right).   
\eea
Here $m_B$ denotes the apparent magnitude and $M$ is the absolute magnitude. In the last equality we have converted between parsecs and megaparsecs, so that we have the right units to describe $H_0$. 

The Pantheon Type Ia supernovae dataset \cite{Scolnic:2017caz} includes observations up to redshift $z = 2.26$, but understandably becomes sparse at higher redshift. In integrating (\ref{ld}), we have a number of options. Since $H(z)$ is analytic, one can simply Taylor expand around $z=0$ and the approximation is reasonable up to $z \approx 0.3$ \footnote{This has the upshot that the results can be explained to high-school students.}. Alternatively, one can employ Pad\'e approximants to increase the radius of convergence, e. g. \cite{Dutta:2018vmq}. The final option is to integrate  (\ref{ld}) numerically and this is the approach we adopt here as it will allow us to identify best-fit parameters for the entire dataset. It does have the downside that the process is more of a black box. 

Before proceeding to the fitting procedure, let us first comment on the absolute magnitude. As touched upon earlier, $H_0$ is degenerate with $M$ and from the Pantheon dataset on its own, one can only determine the (constant) combination $M - 5 \log_{10} H_0$. To remove this degeneracy we will simply assume the nominal value $M = -19.3$ \cite{Hillebrandt:2000ga} \footnote{Alternatively, we could choose $M= - 19.23$, which can be inferred from $H_0$ determined in \cite{Riess:2016jrr}.}. Thus, the value of $H_0$ we obtain will only be valid up to a constant shift, but we will still be able to comment on how it behaves as $z_{\textrm{max}}$ is varied. 

In practice, the data fitting reduces to extremising the following quantity: 
\be
\label{chi}
\chi^2  = \Delta \vec{\mu}^{T} \cdot \mathbf{C}^{-1} \cdot \Delta \vec{\mu}, 
\ee
where $\Delta \vec{\mu} = \vec{\mu} - \vec{\mu}_{\textrm{model}} (H_0, \omega_m)$ is the difference between the data and the model, namely $\Lambda$CDM, which we are assuming only depends on two parameters $(H_0, \omega_m)$. The uncertainty matrix $\mathbf{C}$ can be further decomposed into the statistical matrix $\mathbf{D}_{\textrm{stat}}$ and the systematic covariance matrix $ \mathbf{C}_{\textrm{sys}}$, 
\be
\mathbf{C} = \mathbf{D}_{\textrm{stat}} + \mathbf{C}_{\textrm{sys}}, 
\ee
where the former has only diagonal components. In other words, in the absence of $ \mathbf{C}_{\textrm{sys}}$, extremising this quantity reduces to the usual error-weighted least squares. The inclusion of $\mathbf{C}_{\textrm{sys}}$ allows one to quantify the uncertainties arising from systematics. For practical purposes, as we impose a cut-off $z_{\textrm{max}}$, we will in the process jettison supernovae data at higher redshift, which translates into a cropping of the uncertainty matrix to only include the uncertainties due to supernovae below the cut-off. Note, in order to be more conservative our analysis will include both uncertainties. 

Having set up the problem, let us turn to our first check of the data fitting. At low redshift the Hubble Law should hold, so we can expect to get a reasonable fit for $H_0$ even with a low cut-off $z_{\textrm{max}}$. Here, the actual value of $H_0$ will not be important, just its behaviour with $z_{\textrm{max}}$ will be of interest. As can be seen from Figure 1, the value of $H_0$ returned from the fitting procedure is pretty robust to changes in the cut-off. In particular, it is worth noting that the $1 \, \sigma$ confidence interval is larger at low redshift, where one would expect the uncertainties in the fitting to be large, but it contracts quickly as the cut-off is increased.  Neglecting some wiggles, it should also be clear that there is a straight line corresponding to a constant value of $H_0$ that one can draw through the $1 \, \sigma$ confidence interval.

\begin{figure}[h]
\begin{center}
\includegraphics[width=0.8\textwidth]{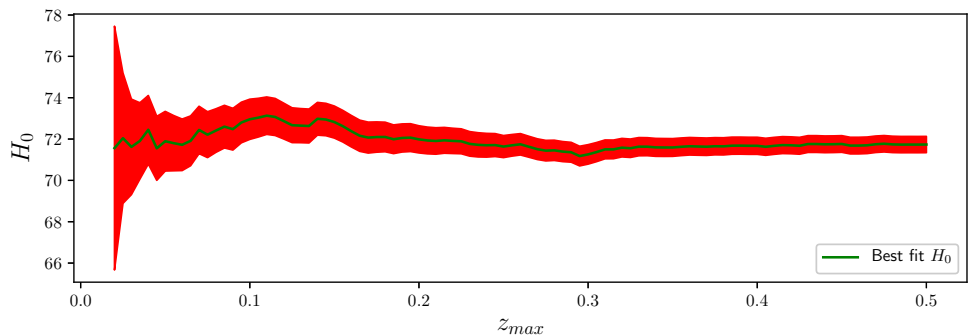}
\end{center}
\vspace{-5mm} 
\caption{Variation in the best-fit value of $H_0$ with $z_{\textrm{max}}$ with $1 \, \sigma$ confidence interval.}
\end{figure}

Now that we are confident that nothing overly suspicious is happening with the best-fit value of $H_0$, let us turn our focus to $\omega_m$ over the same range. From Figure 2 one can see more or less the corresponding plot for $\omega_m$, but we have started fitting from $z_{\textrm{max}} = 0.05$. Bearing in mind that $\omega_m$ is suppressed at low $z$ relative to $H_0$, as is evident from the Taylor expansion, 
\be
H (z) = H_0 \left( 1 + \frac{3}{2} \omega_m z  + \dots \right), 
\ee
it is easy to infer that the uncertainties in fitting $\omega_m$ will be greater from the outset at low $z$. This is evident from Figure 2 and for this reason we have cut the lower redshift best-fits. Again, we observe that as $z_{\textrm{max}}$ becomes larger, the $1 \, \sigma$ confidence interval contracts and the best-fit value converges to a well-defined constant value within this interval. For comparison we have added the lower bound on the Planck value and beyond {$z_{\textrm{max}} \approx 0.16$}, we see there is good agreement. 

However, within the range {$0.1 \leq z_{\textrm{max}} \leq 0.16$, there is a perceivable} departure from the Planck value. First, the best-fit value of $\omega_m$ is negative in places, but this is within $1 \, \sigma$ of zero. Bearing in mind that $\omega_m$ is an energy density, this could be pointing towards an unphysical result in a given range. One could of course bound $\omega_m$, but the extremization of (\ref{chi}) is a well-defined math problem and a computer is agnostic to the meaning of $\omega_m$, so we should try to live with the result. 

\begin{figure}[h]
\begin{center}
\includegraphics[width=0.8\textwidth]{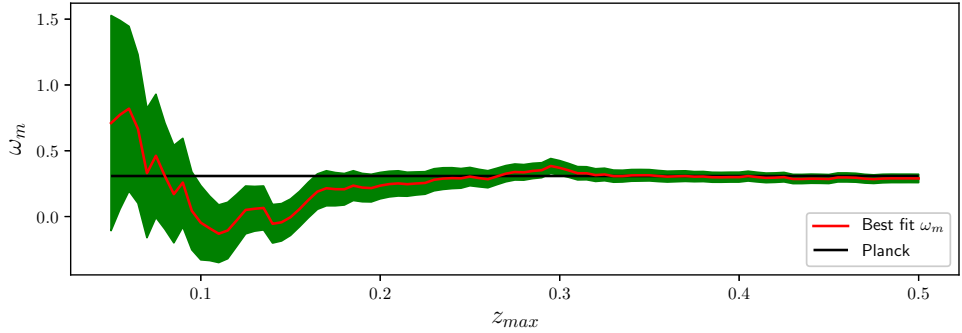}
\end{center} 
\vspace{-5mm} 
\caption{Variation in the best-fit value of $\omega_m$ with $z_{\textrm{max}}$ with $1 \, \sigma$ confidence interval.}
\end{figure}

At this stage it makes sense to drill down a bit more into the figures and the estimation of confidence intervals. To this end, we make use of the lmfit Python package to estimate the confidence intervals for a selection of best-fit values of $\omega_m$ with varying $z_{\textrm{max}}$. The result is presented in Table 1, {where we have quoted $1 \, \sigma$ ($68.27 \%$) and $2 \, \sigma$ ($95.45\%$) confidence intervals}. The first thing to note is that when the entire dataset is considered at $z_{\textrm{max}} = 2.3$, our result agrees with the quoted Pantheon result {$\omega_m = 0.298 \pm 0.022$} \cite{Scolnic:2017caz}, thereby providing an important consistency check on our fitting. 

\begin{figure}[h]
\begin{center}
\includegraphics[width=0.6\textwidth]{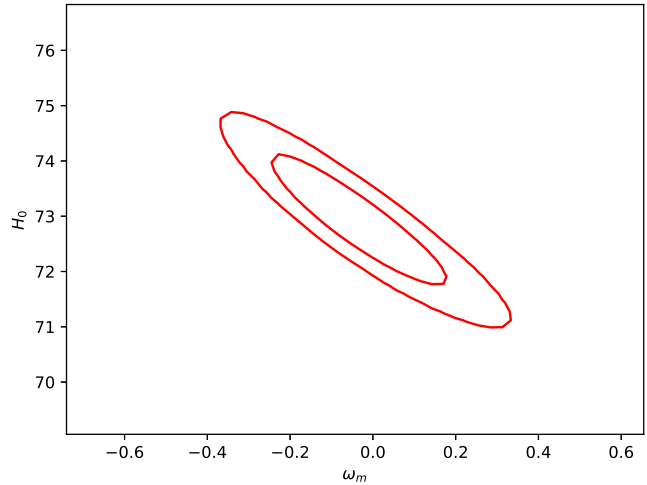}
\end{center} 
\vspace{-5mm} 
\caption{$1$ and $2 \, \sigma$ confidence intervals for best-fit values at $z_{\textrm{max}} = 0.147$ using lmfit.}
\end{figure}

{Moreover, we note that beyond $z_{\textrm{max}} \approx 0.16$ we see little deviation from the Planck value. However, as is clear from Table 1, in a window between $z_{\textrm{max}} = 0.1$ and $z_{\textrm{max}} = 0.16$, we notice a discrepancy in excess of 1 $\sigma$, which exceeds 2 $\sigma$ around $z_{\textrm{max}} \approx 0.145$. In support of this claim, we present the $z_{\textrm{max}}  = 0.147$ confidence ellipses in Figure 3. As is clear from Figure 2 and Table 1, this is not a discrepancy at an isolated value of redshift, but one within a range of redshifts. This is clearly some feature in the Pantheon dataset. }

\begin{table}[h]
\begin{center}
\begin{tabular}{c||ccccc}
$z_{max}$ & $95.45$\% & $68.27$\% & Best-fit &  $68.27$\% & $95.45$\% \\
\hline
\hline
$0.1$ & $-0.546$ & $-0.282$ & $-0.047$ &  $+0.304$ & $+0.634$ \\
$0.12$ & $-0.374$ & $-0.192$ & $-0.029$ &  $+0.205$ & $+0.426$ \\
$0.14$ & $-0.280$ & $-0.144$ & $-0.055$ &  $+0.152$ & $+0.314$ \\
$0.147$ & $-0.266$ & $-0.137$ & $-0.043$ &  $+0.145$ & $+0.299$ \\
$0.16$ & $-0.255$ & $-0.131$ & $0.126$ &  $+0.139$ & $+0.286$ \\
$0.2$ & $-0.193$ & $-0.099$ & $0.235$ &  $+0.105$ & $+0.215$ \\
$0.5$ & $-0.062$ & $-0.032$ & $0.290$ &  $+0.033$ & $+0.067$ \\
$1$ &  $-0.044$ & $-0.022$ & $0.294$ &  $+0.023$ & $+0.047$  \\
$2.3$ & $-0.041$ & $-0.021$ & $0.298$ &  $+0.022$ & $+0.044$  \\
\end{tabular}
\end{center} 
\caption{Best-fit values of $\omega_m$ and confidence intervals for given $z_{\textrm{max}}$.}
\end{table}

\section{Discussion}
In this note we have studied the Pantheon Type Ia supernovae dataset by performing fits of the $\Lambda$CDM cosmological model using a running cut-off $z_{\textrm{max}}$. In line with expectations, we have seen that $H_0$, or at least the quantity from which $H_0$ can be determined given $M$, varies little with $z_{\textrm{max}}$. Shifting our focus to the matter density $\omega_m$, we have a number of observations. First, we recover the Pantheon result over the entire dataset, thus validating our methods. We also observe that attempts to determine $\omega_m$ at low redshift are inconclusive, and bearing in mind that it is suppressed relative to $H_0$ in $z$, this is an understandable result. What is intriguing is the dip in the best-fit value of $\omega_m$ between $z_{\textrm{max}} \approx 0.1$ and {$z_{\textrm{max}} \approx 0.16$}. Recalling that (\ref{LCDM}) is a function and that the parameters $(H_0, \omega_m)$ for \textit{any range of redshift} should ideally be constant within a $1 \, \sigma$ confidence window, this is a surprising result. Moreover, one would expect any discrepancy between the best-fit value of $\omega_m$ and the Planck value to be dressed by a $1 \, \sigma$ confidence interval, but we have found this not to be the case. This points to a potential tension in $\omega_m$. 

{Neglecting some statistical fluke,} the simplest resolution to the discrepancy with the Planck value is that we may have underestimated the uncertainties, but as we recover the Pantheon result, this is not obviously the case. Another possibility is that there is some inconsistency between data points in the dataset, which is not evident when one imposes a high redshift cut-off. So, instead of testing $\Lambda$CDM, we may in fact be testing the Pantheon dataset. On the flip side, if the data holds up, and we will know going forward as future experiments increase the size of the Type Ia supernovae sample \cite{Scolnic:2019apa}, {the perceived matter underdensity could be an indication of a local cosmic void}, or more generally our analysis could suggest that $\Lambda$CDM is breaking down at low redshift. 

{In the case of the former, it is interesting to note that galaxy surveys have reported underdensities in the range $z < 0.7$ \cite{Carrick:2015xza, Whitbourn:2013mwa,Keenan:2013mfa}, but these fall outside of the $z = 0.1$ to $z = 0.16$ window that interests us. Nevertheless, one can in principle model a cosmic void, but it has been shown that it has a negligible effect on the Hubble constant \cite{Kenworthy:2019qwq}. It would be interesting to apply similar analysis to see if the difference with the Planck value can be accounted for by a cosmic void.}

If the discrepancy cannot be explained by mundane explanations, it is tempting to speculate that we are looking at new physics. Although the discrepancy is not so significant, it is on par with current tension in cosmic shear $S_8 \equiv \sigma_8 ( \omega_m/0.3)^{\alpha}$ \cite{Heymans:2012gg, Troxel:2017xyo, Joudaki:2017zdt, Hikage:2018qbn}. Interestingly, the latter may also be pointing to a lower value of $\omega_m$ relative to Planck. In principle, such a feature, if real, could be explained by a coupling between dark matter and dark energy, whereby dark matter becomes dark energy at late times. A  coupling of this nature can be expected to manifest itself in a potential underdensity in matter and an increase in the Hubble constant at late times.
 
\section*{Acknowledgements}
We thank S. Brahma,  M. Bureau, W. Hossain, D. Spergel, P. Steinhardt and A. Trautner for comments and discussion. This work was initiated with Maurice van Putten, who we thank for sharing his insights. We also thank L. Macri \& D. Scolnic for patient explanations and wish to acknowledge the benefit of attending the APCTP workshop ``$H_0$ tension \& Swampland: theory confronts reality". We thank M. M. Sheikh-Jabbari \& H. Yavartanoo for comments on earlier drafts. 

\appendix
\section{MCMC}
Since curve fitting via Python and the lmfit package is a little opaque, one can try to get a better feel for the data by employing Markov Chain Monte Carlo (MCMC). To this end, in this section we revisit the task of identifying the best-fit values of $(H_0, \omega_m)$ at $z_{\textrm{max}} = 0.147$. The goal is to recover the central values, the covariance matrix and the confidence interval ellipses from a simple Metropolis-Hastings algorithm. 

In practice, we avoid a burn-in phase by simply starting the Markov Chain from the best-fit values returned by the fitting procedure. This is justifiable since we are more interested in estimating the error and producing confidence intervals. We assume the probability is $P \propto \exp ( - \chi^2/2)$, where the normalisation factor does not interest us (we are only interested in relative probabilities) and $\chi^2$ is defined in (\ref{chi}). We explore parameter space by picking $H_0$ and $\omega_m$ from normal distributions with standard deviations, $\sigma = 1.5$ and $\sigma = 0.1$, respectively, which allows us to adequately explore the parameter space. Below we present the results of 100,000 iterations with a final acceptance rate of approximately 22\%. The generated configurations are presented in Figure 4 and it can be confirmed that the parameter space is being well explored. 

\begin{figure}[h]
\begin{center}
\includegraphics[width=0.8 \textwidth]{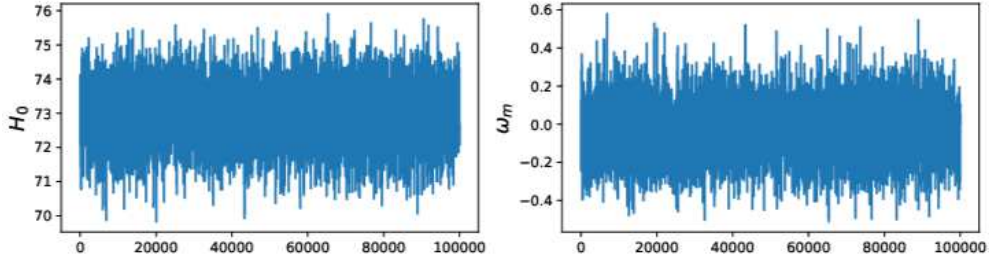}
\end{center}
\vspace{-5mm} 
\caption{The trace plots for parameters $H_0$ and $\omega_m$ plotted against the number of iterations.}
\end{figure}

It is instructive to compare the best-fit values and covariance matrices returned by curve fitting and MCMC. To four decimal places the best-fit values are 
\be
(H_0, \omega_m)_{\textrm{fitting}} =  (72.9419, - 0.0429), \quad  (H_0, \omega_m)_{\textrm{MCMC}} =  (72.9053, - 0.0319), 
\ee
while the returned covariance matrices $C$ are  
\be
C_{\textrm{fitting}} = \left( \begin{array}{cc} 0.0195 & - 0.0992 \\ - 0.0992 & 0.6073 \end{array} \right), \quad  C_{\textrm{MCMC}} = \left( \begin{array}{cc}  0.0198 & - 0.1006 \\ - 0.1006 & 0.6178 \end{array} \right). 
\ee

\begin{figure}[h]
\begin{center}
\includegraphics[width=0.5 \textwidth]{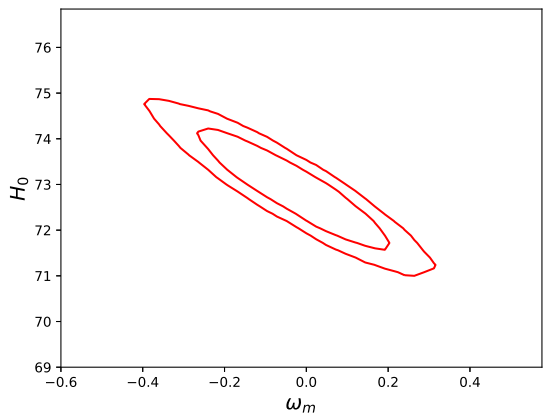}
\end{center} 
\vspace{-5mm} 
\caption{$1$ and $2 \, \sigma$ confidence intervals for best-fit values at $z_{\textrm{max}} = 0.147$ using MCMC.}
\end{figure}

Although there is a slight discrepancy in $\omega_m$, it is clear that the covariance matrices show excellent agreement. With the covariance matrix in hand, it is easy to generate the corresponding bivariate normal distribution and identify the 1 and 2 $\sigma$ ellipses. We present the result in figure 5, and as expected, it shows good agreement with Figure 3. Ultimately, since the covariance matrices are almost identical, this outcome is no surprise.

\end{document}